\documentclass[twocolumn]{aastex631}

\shorttitle{Giant radio quasars: composite optical spectra}
\shortauthors{Ku\'zmicz et al.}
\graphicspath{{./}{figures/}}

\begin{document}

\title{Giant radio quasars: composite optical spectra}

\author[0000-0002-3097-5605]{Agnieszka Ku\'zmicz}
\affiliation{Astronomical Observatory, Jagiellonian University, ul. Orla 171, 30-244 Krakow, Poland}
\affiliation{Queen Jadwiga Astronomical Observatory in Rzepiennik Biskupi, 33-163 Rzepiennik Strzy\.zewski, Poland}

\author[0000-0001-8561-4228]{Sagar Sethi}
\affiliation{Astronomical Observatory, Jagiellonian University, ul. Orla 171, 30-244 Krakow, Poland}

\author[0000-0002-0870-7778]{Marek Jamrozy}
\affiliation{Astronomical Observatory, Jagiellonian University, ul. Orla 171, 30-244 Krakow, Poland}



\begin{abstract}

We present the composite optical spectrum for the largest sample of giant radio quasars (GRQs). They represent a rare subclass of radio quasars due to their large projected linear sizes of radio structures, which exceed 0.7 Mpc. To construct the composite spectrum, we combined 216 GRQ's optical spectra from Sloan Digital Sky Survey (SDSS). As a result, we obtained the composite spectrum covering the wavelength range from 1400 {\AA} to 7000 {\AA}. We calculated the power-law spectral slope for GRQ's composite, obtaining $\alpha_{\lambda}=-1.25$ and compared it with that of the smaller-sized radio quasars, as well as with the quasar composite spectrum obtained for large sample of SDSS quasars. We obtained that the GRQ's continuum is flatter (redder) than the continuum of comparison quasar samples. We also show that the continuum slope depends on core and total radio luminosity at 1.4 GHz, being steeper for higher radio luminosity bin. Moreover, we found the flattening of the continuum with an increase of the projected linear size of radio quasar. We show that $\alpha_{\lambda}$ is orientation-dependent, being steeper for a higher radio core-to-lobe flux density ratio which is consistent with AGN unified model predictions. For two GRQs, we fit the spectral energy distribution using X-CIGALE code to compare the consistency of results obtained in the optical part of the electromagnetic spectrum with broad-band emission. The parameters obtained from the SED fitting confirmed the larger dust luminosity for the redder optical continuum.

\end{abstract}

\keywords{galaxies: active -- galaxies: nuclei -- galaxies: structure}


\section{Introduction} \label{sec:intro}

Quasars are the most luminous active galactic nuclei (AGNs) which, unlike galaxies, emit a strong nonstellar continuum over the full range of the electromagnetic spectrum. The broadband overall continuum (spectral energy distribution; SED) of quasars is relatively flat over a wide range of frequencies (e.g. \citealt{carleton1987}, \citealt{elvis1994}) having two characteristic components: the Big Blue Bump and the infrared bump. The Big Blue Bump extends from near-infrared at 1 $\mu$m to past 1000 {\AA} in the UV part of the spectrum where a substantial part of AGN bolometric luminosity is emitted. The UV/optical continuum is explained as the thermal emission from an accretion disc which is parametrized by a power-law relation $f_{\nu}\sim \nu^{\alpha}$, where $\alpha$ is the continuum slope. It is assumed that the accretion disk is geometrically thin and optically thick \citep{shakura1973}, where each annulus of the accretion disc radiates as a black body with a characteristic temperature of a milion Kelvins. The infrared bump extends from $\sim$100 to $\sim$ 1 $\mu$m. It is believed that it arises from the re-emitting of the UV/optical emission by the dust surrounding the central AGN. 

The optical/UV part of quasar spectra are dominated by a power-law continuum and a series of broad emission lines. Despite detailed differences in individual quasar spectra, they are quite similar in general. To study the overall properties of quasars, the composite spectra can be used. They can be obtained by combining large number of quasar spectra. The resultant composite spectrum has a high signal-to-noise ratio, therefore both the intrinsic continuum shape, as well as weak emission features not detectable in individual spectra can be studied. In the literature, composite spectra were obtained for different quasar samples. For example, \cite{zheng1997} obtained composite spectrum for 284 quasars observed by the Hubble Space Telescope, \cite{brotherton2001} for 657 quasars catalogued in the FIRST Bright Quasar Survey, \cite{vanden2001} for 2204 quasars from SDSS covering a wavelength range of 800 -- 8555~{\AA} in the rest frame, and \cite{harris2016} for 102150 quasars with redshift $2.1\leq z \leq 3.5$ from the Baryon Oscillation Spectroscopic Survey.  

The composite quasar spectra are very useful in studies of various scientific problems. For example, they were used by \cite{vanden2001} for measurement of emission line shifts relative to laboratory wavelengths and to find undiscovered emission lines which can be detectable in high signal-to-noise composites (see also \citealt{francis1991}). They were also used for testing the accretion disk models (e.g., \citealt{koratkar1999}) and for investigations of spectral parameters evolution with redshift \citep{xie2015, carballo1999}. Composite spectra were also used to study the differences between radio-loud and radio-quiet quasars (\citealt{cheng1991, zheng1997}), where the authors obtained that the spectra of radio-loud quasars are steeper than the spectra of radio-quiet ones. Furthermore, the studies of \cite{shankar2016} focused on the connection between accretion parameters (i.e. black hole mass, accretion rate, and black hole spin) and the shape of the spectral continuum. 

Giant radio quasars (GRQs) are a rare class of AGNs. The projected linear size of their radio structures exceed 0.7 Mpc. Although the number of radio sources with such large size still grows up, we know only $\sim$ 300 GRQs \citep{kuzmicz2021, dabhade2020}. Most of them were found mainly in the northern hemisphere thanks to a large radio and optical surveys like NRAO VLA Sky Survey (NVSS; \citealt{condon1998}), Faint Images of the Radio Sky at Twenty-Centimeters (FIRST; \citealt{becker1995}), low-frequency LOFAR Two-metre Sky Survey first data release \citep[LoTSS;][]{shimwell2017}, Sloan Digital Sky Survey (SDSS; \citealt{abolfathi2018}). Despite many studies aimed at finding the explanation of GRQ growth to Mpc size (e.g. \citealt{mack1998, ishwara1999, machalski2006, subrahmanyan2008, machalski2009, machalski2011, kuzmicz2012, kuzmicz2019, dabhade2020, kuzmicz2021}) the processes responsible for their origin are not fully understood.  

According to the commonly accepted model \citep{blandford1977, meier2001}, radio jets are produced by spinning black holes. It is believed that the processes responsible for radio jet emission are closely related to the accretion processes which manifest in the optical/UV spectrum as a Big Blue Bump.
 
In this study, we want to investigate the overall properties of GRQs using their composite spectrum. Thus, we want to compare it with other composites obtained for different quasar samples and to see if they show any specific features which can be characteristic for this class of AGNs. 
  
Throughout the paper, we use the following values for cosmological parameters: H$_0$=71 km $\rm s^{-1} Mpc^{-1}$, $\Omega_M$=0.27, $\Omega_{\Lambda}$=0.73.

\section{Quasar sample}

In our analysis, we used the largest sample of GRQs from \cite{kuzmicz2021} where the authors collected 272 quasars in the redshift range of 0.1 $<$ z $<$ 3 and with a median value of projected linear size D=0.9 Mpc. Most of the quasars in the sample ($\sim$87\%) were found by cross-matching NVSS and FIRST radio maps with PanSTARS \citep{flewelling2020} and SDSS optical images. In our analysis, we used 237 quasars from the entire sample of GRQs for which the SDSS optical spectra were available. 

As a comparison sample, we used the sample of smaller radio quasars (SRQ) from \cite{kuzmicz2021}. It contains 367 quasars with projected linear size between 0.2 and 0.7 Mpc and it covers a similar redshift as the GRQ sample.

Both GRQ and SRQ samples are not complete but are homogenous in terms of redshift coverage, total radio luminosity, and core radio luminosity values. 

\section{Composite spectra}
\label{composite}

To obtain the GRQ and SRQ composite spectra, we corrected each quasar spectrum for Galactic extinction and shifted to the rest frame using the redshift value given in the SDSS. Then, each spectrum was re-binned into a dispersion of 1 {\AA} per bin conserving flux and normalized to the unit average flux density over the wavelength interval of 3600-3700 {\AA} (in the rest frame). In this wavelength range, there are no strong emission lines and the contamination by Fe emission is not prominent. The spectra of some quasars from our samples do not cover this wavelength range (mainly high redshift QSOs), therefore to obtain composite spectra we use only 216 GRQs and 328 SRQs for which normalization was possible. To study the properties of large-scale continuum shape, we adopted the geometric mean to obtain the composite spectrum, which prevents the global shape of the quasar power-law continuum \citep{vanden2001}.   

\begin{figure}[h!] 
\centering 
\includegraphics[width=1\columnwidth]{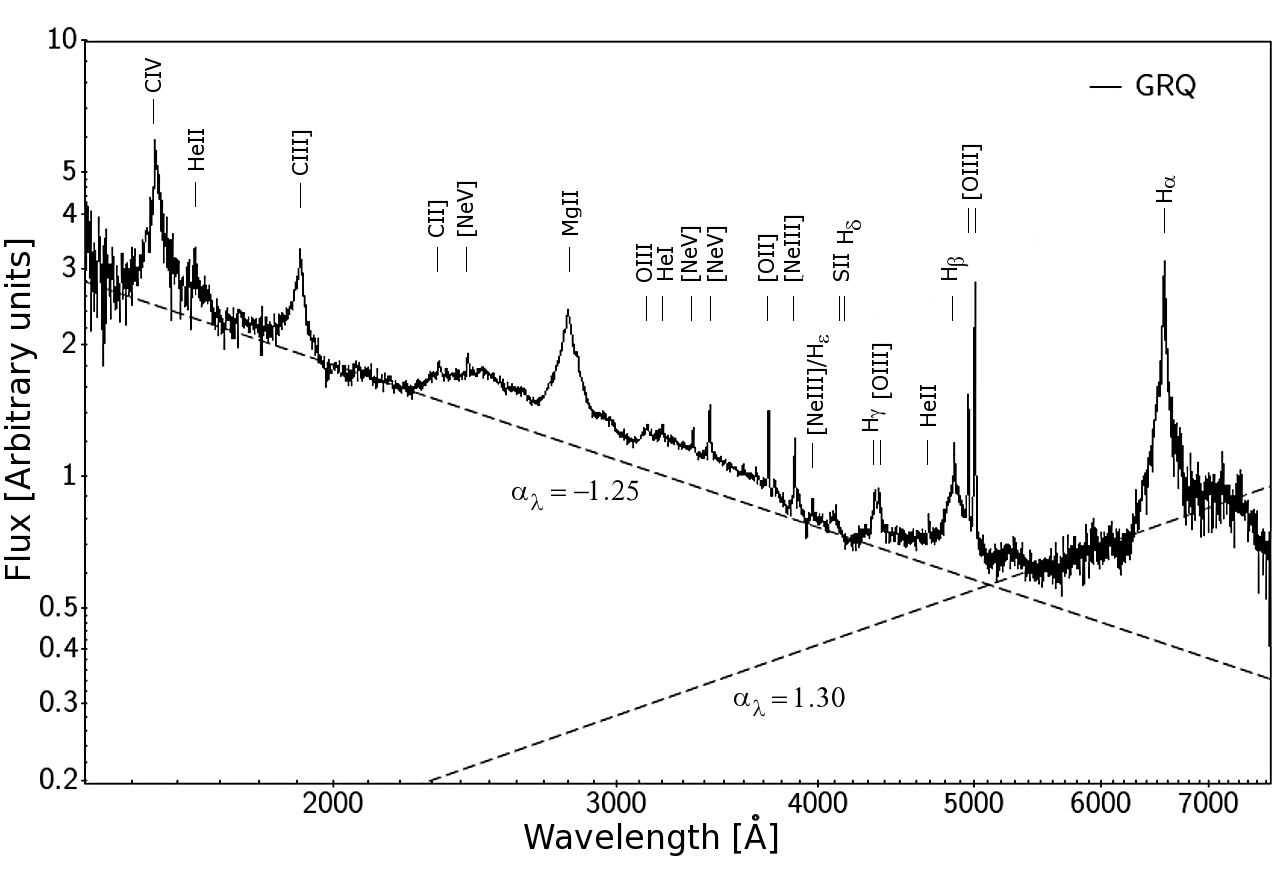}\\ 
\includegraphics[width=1.01\columnwidth]{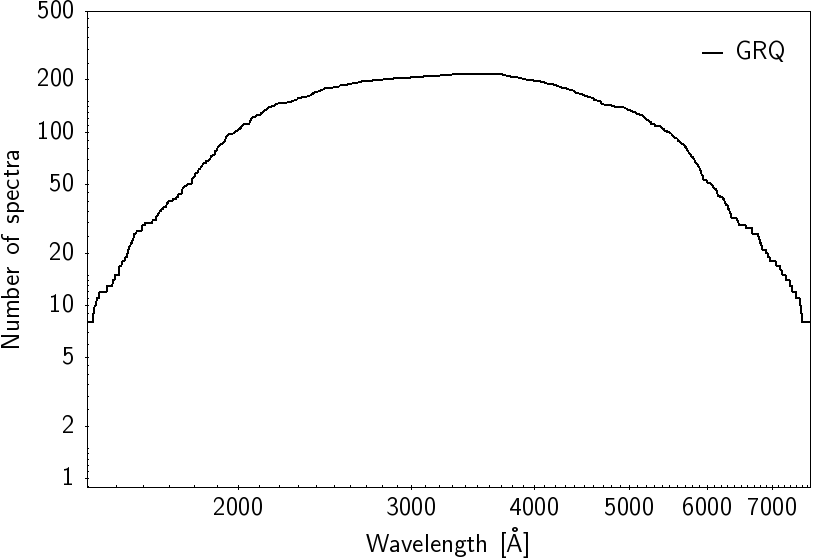}\\ 
\caption{{\bf Top:} composite GRQ spectrum with labeled emission line names. Power-law fits are plotted as dashed lines. {\bf Bottom:} number of combined GRQ's spectra.} 
\label{grq} 
\end{figure}

\begin{figure}[h!] 
\centering 
\includegraphics[width=1\columnwidth]{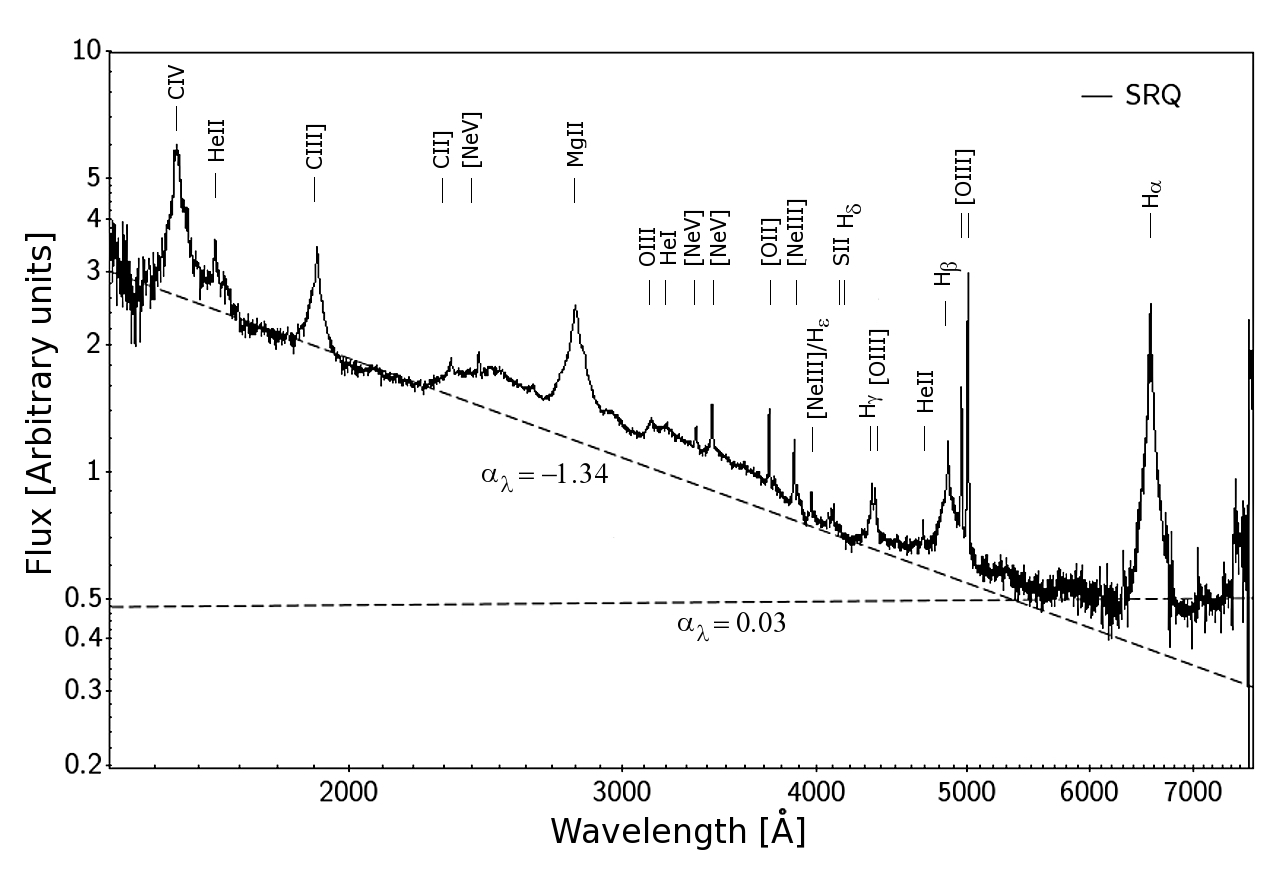}\\ 
\includegraphics[width=1.01\columnwidth]{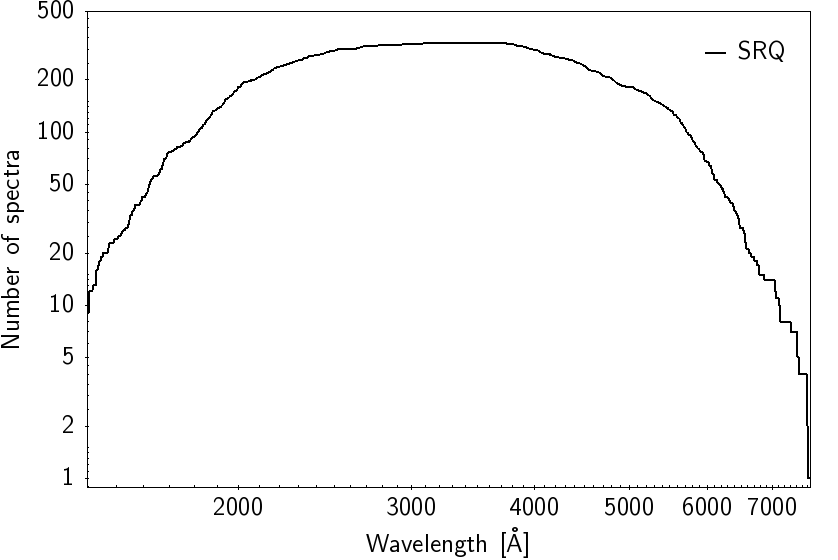}\\ 
\caption{{\bf Top:} composite SRQ spectrum with labeled emission line names. Power-law fits are plotted as dashed lines. {\bf Bottom:} number of combined SRQ's spectra.} 
\label{srq} 
\end{figure}

The composite spectra for GRQs and SRQs are presented in Figures \ref{grq} and \ref{srq} on a log-log plot. In the bottom panels we also plot the number of quasar spectra that contribute to each 1~{\AA} bin for both samples. We fitted the power-law continuum to the composite spectra which appears as a straight line on a log-log plot. For a fit we used the small wavelength regions 2210-2230~{\AA} and 4200-4230~{\AA} where the contamination by emission lines is expected to be very weak. We obtained the slopes of power-law $\alpha_{\lambda}=-1.25\pm0.01 $ and $\alpha_{\lambda}=-1.34\pm0.01$ for GRQs and SRQs, respectively. The uncertainties of power-law slopes are determined as the standard deviation of a linear least-squares fit. However, the obtained uncertainties can be understated due to the small wavelength regions used for power-law fit. \\ 

In Figures \ref{grq} and \ref{srq} it is visible that the fitted power laws do not fit well in the red part of the spectrum (beyond 5000~{\AA}). It can be a result of contamination by the stellar light from the host galaxy but also can be caused by an internal change in the quasar continuum emission \citep{vanden2001}. Therefore, in the red part of GRQ and SRQ composites, we fitted separate power laws in the wavelength ranges of 6005-6035~{\AA} and 7160-7180~{\AA}. We obtained slope $\alpha_{\lambda}=1.30\pm0.07$ for GRQs and $\alpha_{\lambda}=0.03\pm0.08$ for SRQs. The slopes obtained in the blue and red part of the spectrum slightly differ from the slopes obtained by \cite{vanden2001}. They received $\alpha_{\lambda}=-1.56$ below 5000~{\AA} and $\alpha_{\lambda}=0.45$ beyond 6000~{\AA} for the sample of 2204 SDSS quasars in a redshift range $0.04<z<4.79$. The sample of quasars used by \cite{vanden2001} was selected based on SDSS magnitude limits as well as, for high redshift sources, also on FIRST radio core emission. Therefore, it can be a good reference sample for specific classes of quasars. In general, extended radio quasars, i.e., GRQs and SRQs, have a flatter (less negative value of $\alpha_{\lambda}$) blue part of the optical spectrum compared to quasars from \cite{vanden2001}. In the red part of the spectrum, GRQs have the highest $\alpha_{\lambda}$ value, but as it is visible in the bottom panel of Figure \ref{grq}, there is a very low number of quasar spectra ($\sim$10) contributing to the final composite spectrum beyond 7160~{\AA}. As we wrote above, the 7160-7180~{\AA} wavelength range was used for power-law fitting, therefore the obtained slope value for the red part of the spectrum may be not representative for the GRQ and SRQ samples.\\ 

Comparing our result to other studies, the continuum slopes below 5000~{\AA} for GRQ and SRQ samples are also higher than the slopes obtained by other authors, e.g., \cite{brotherton2001} received $\alpha_{\lambda}=-1.54$ for the FIRST Bright Quasar Survey sample, \cite{shankar2016} obtained $\alpha_{\lambda}$ below -1.5 for SDSS DR7 quasars. It indicates that extended radio quasars have in general redder spectra than other quasar samples. This effect is related rather to the intrinsic reddening of quasars (e.g., \citealt{richards2003}) than to sample selection effects (e.g., higher redshift samples have redder spectral slopes; \citealt{francis1993, xie2016}). Assuming that the intrinsic extinction is described by the Small Magellanic Cloud (SMC) extinction curve \citep{gordon1998}, the difference in $\alpha_{\lambda}$ values for GRQs and that obtained by, e.g. \cite{brotherton2001} corresponds to $E(B-V)\sim0.07$.
However, it was shown by \cite{baker1995} that the redder spectra are observed in lobe dominated quasars indicating the larger dust obscuration in quasars viewed at larger inclinations. Due to the fact that very extended radio sources can be easier to identifify when their inclination angles are larger (their projected angular size on the sky plane is larger) the redder spectra in GRQs and SRQs can largely be a result of projection effects.

\section{Continuum slope - radio luminosity dependence} 
\label{P} 

\begin{figure} 
\centering 
\includegraphics[width=1\columnwidth]{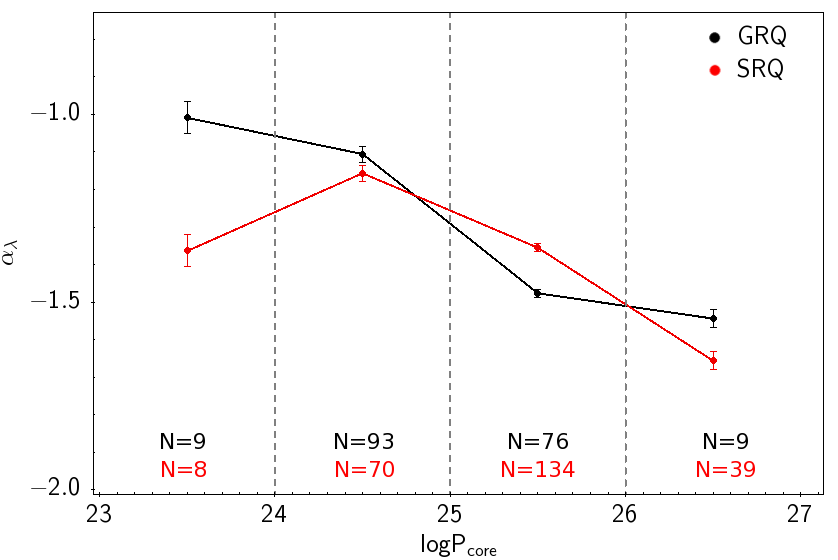}\\ 
\includegraphics[width=1\columnwidth]{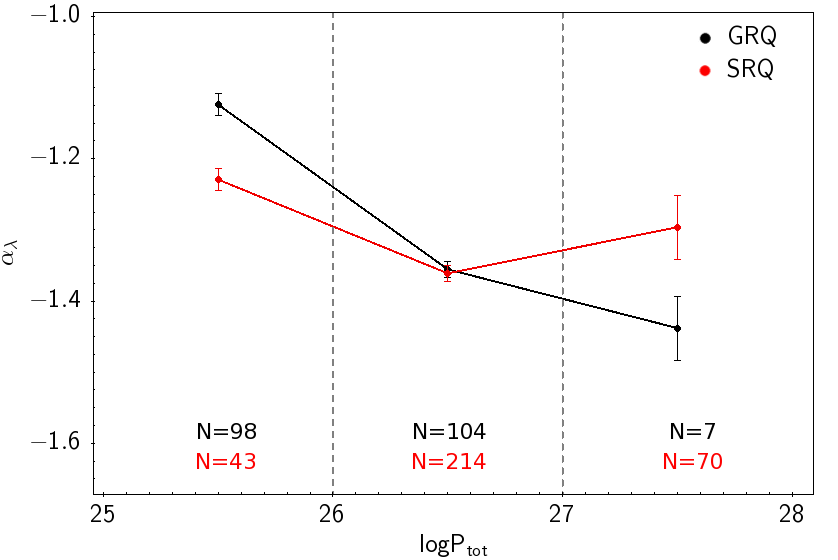}\\ 
\caption{Power-law slope vs. core radio luminosity (top) and total radio luminosity (bottom) at 1.4 GHz for GRQ and SRQ samples. By vertical dotted lines, we marked the widths of radio luminosity bins in which we compute composite spectra. For each radio luminosity bin, we wrote the number of spectra used to the composite spectrum calculation; in black for GRQs and in red for SRQs.} 
\label{P_slope} 
\end{figure}

For GRQ and SRQ samples we computed the composite spectra for different radio luminosity bins using the method described in Section \ref{composite}. We separately considered the radio core (P$_{\rm core}$) and total radio (P$_{\rm tot}$) luminosities at 1.4 GHz which were measured by \cite{kuzmicz2021}. The obtained continuum slopes for each radio luminosity bin are listed in Table \ref{tab1} and they are plotted in Figure \ref{P_slope}, where we also give the number of spectra contributing to the final composite spectrum. It is visible that there is a tendency to steepen the continuum slope for higher core and total radio luminosities.  

\begin{table}[h!] 
\centering 
\caption{Slopes of composite continuum for different P$_{\rm core}$ and P$_{\rm tot}$ at 1.4 GHz for GRQ and SRQ samples.} 

\label{tab1} 
\begin{tabular}{ccc} 
\hline 
bin			& $\alpha_{\lambda}$ for GRQ & $\alpha_{\lambda}$ for SRQ \\ 
\hline \hline 
$\log{P_{\rm core}}<24$		& $-$1.01$\pm$0.04	& $-$1.36$\pm$0.04\\ 
$24<\log{P_{\rm core}}<25$	& $-$1.11$\pm$0.02	& $-$1.16$\pm$0.02\\ 
$25<\log{P_{\rm core}}<26$	& $-$1.48$\pm$0.02	& $-$1.36$\pm$0.01\\ 
$\log{P_{\rm core}}>26$		& $-$1.54$\pm$0.04	& $-$1.66$\pm$0.02\\ 
\hline 
$25<\log{P_{\rm tot}}<26$	& $-$1.12$\pm$0.02	& $-$1.23$\pm$0.02\\ 
$26<\log{P_{\rm tot}}<27$	& $-$1.36$\pm$0.01	& $-$1.36$\pm$0.01\\ 
$\log{P_{\rm tot}}>27$		& $-$1.44$\pm$0.04	& $-$1.30$\pm$0.02\\ 
\hline 
\end{tabular} 
\end{table} 

A similar dependence was earlier obtained by \cite{carballo1999}. The authors divided quasars into high-power and low-power quasars using the division limit of radio luminosity at 408 MHz equal to 10$^{34.85}$ erg s$^{-1}$ Hz$^{-1}$. They obtained continuum slopes equal to $\alpha_{\lambda}=-1.29$ and $\alpha_{\lambda}=-1.49$ for lower and higher radio luminosity bins below $\lambda$=2600 {\AA} (originally \citealt{carballo1999} use frequency index $\alpha_{\nu}$ which is related to $\alpha_{\lambda}$ by $\alpha_{\lambda}=-(\alpha_{\nu}+2)$; \citealt{vanden2001}).  

Our result seems to be in agreement with theoretical predictions. According to commonly accepted models (e.g., \citealt{blandford1977}) radio jets are produced by rapidly spinning black holes. It may be expected that in GRQs, the mechanism responsible for radio jet emission should be very efficient to fuel and collimate radio jets over hundreds of kiloparsecs. It was found by \cite{schulze2017} that radio-loud quasars have systematically higher black hole spins, therefore it is expected that they should have a bluer (steeper) UV/optical continuum (according to \citealt{davis2011}). However, \cite{shankar2016} did not find clear evidence of such dependence for a large sample of radio-loud and radio-quiet quasars, obtaining even opposite results. Moreover, in the magnetohydrodynamical simulations of accretion discs provided by \cite{moscibrodzka2016} the authors obtained that the black hole spin may not be the main driver of radio jet emission, hence the radio luminosity does not change as a function of black hole spin. Based on earlier studies (e.g., \citealt{kuzmicz2012}, \citealt{dabhade2020}), we know that the properties of central AGN in GRQs, like black hole mass, accretion rate, and black hole spin are very similar to those obtained for other radio sources, therefore the dependence between $\alpha_{\lambda}$ and radio luminosity may have a different origin.  

\begin{figure} 
\centering 
\includegraphics[width=1\columnwidth]{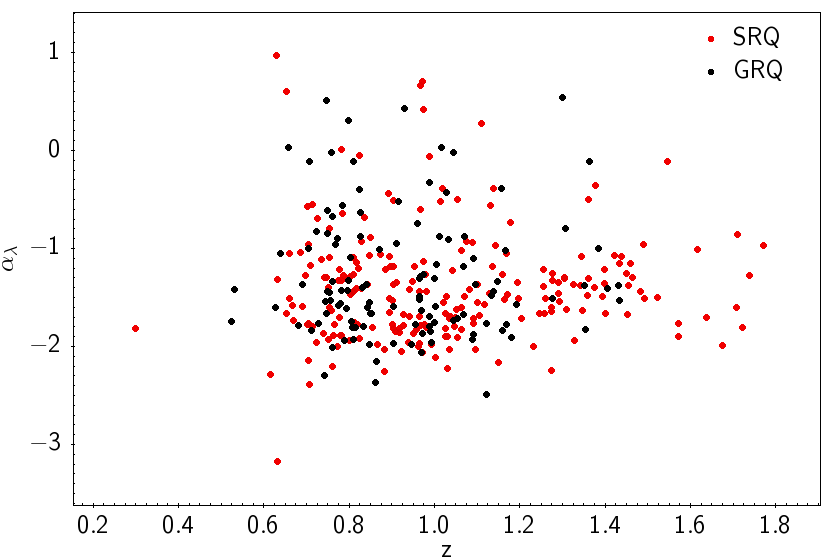}\\ 
\caption{Continuum slope - redshift dependence for GRQs and SRQs.} 
\label{zslope} 
\end{figure}

On the other hand, the visible $\alpha_{\lambda}$-radio luminosity dependence may be a result of continuum slope evolution with redshift. It was found, e.g., by \cite{xie2015} who study the quasars up to z$\gtrsim$3, that the quasar continuum is redder (has less negative $\alpha_{\lambda}$) toward higher redshift, due to cosmic dust extinction. We checked if for our samples of GRQs and SRQs the slope--redshift dependence is significant. In Figure \ref{zslope} we plot the $\alpha_{\lambda}$ for individual quasars as a function of redshift. It is visible that there is a very weak dependence between $\alpha_{\lambda}$ and z for both samples, with a negligible tendency of $\alpha_{\lambda}$ increase with redshift. It also should be pointed out that all quasars from our samples (for which we were able to measure $\alpha_{\lambda}$) have relatively low redshift ($0.3<z<1.8$), therefore the $\alpha_{\lambda}$--z evolution for our samples of quasars may be not prominent in such a narrow band of redshift. 

\begin{figure}[h!] 
\centering 
\includegraphics[width=1\columnwidth]{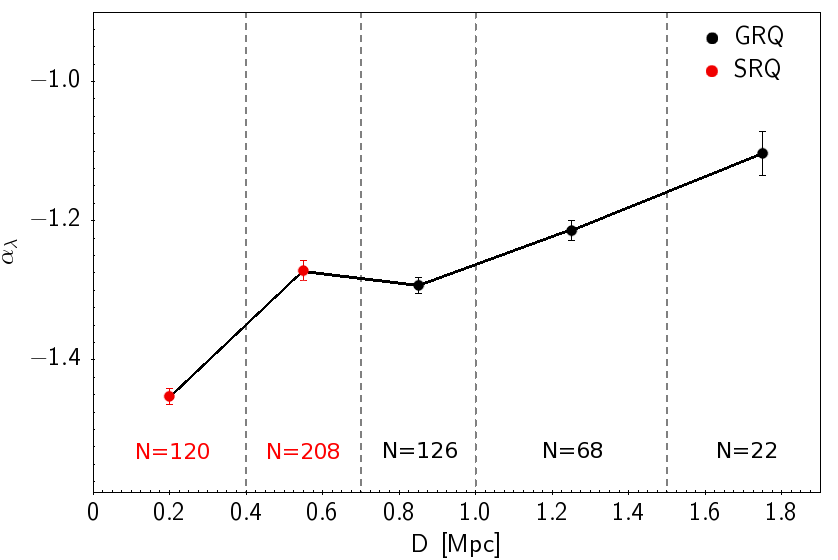} 
\caption{The power law slope vs. projected linear size. Description as in Figure \ref{P_slope}.} 
\label{D_slope} 
\end{figure}

For the samples of GRQs and SRQs, we also checked the dependence between $\alpha_{\lambda}$ and the projected linear size. We binned the quasar spectra according to size values and computed the composite spectrum in each bin. The fitted continuum slopes are listed in Table \ref{tab2}. In Figure \ref{D_slope} it is clearly visible that the larger radio structure of the quasar is, the optical continuum becomes flatter. According to our knowledge about giant radio sources, their large size may be a consequence of their advanced age. Therefore, the dependence visible in Figure \ref{D_slope} can also be interpreted as the continuum slope evolution with radio source age. However, this effect may be also a result of the continuum slope flattening with decreasing total radio luminosity visible in Figure \ref{P_slope}. According to the radio source evolutionary models by \cite{kaiser1997}, the P$_{\rm tot}$ decreases as the projected linear size increases. This dependence is clearly visible in a luminosity - linear size (P--D) diagram (e.g., \citealt{kuzmicz2021} for GRQs and SRQs). 

\begin{table}[h!] 
\centering 
\caption{Slopes of composite continuum for different projected linear size bins.} 
\label{tab2} 
\begin{tabular}{cc} 
\hline 
bin			& $\alpha_{\lambda}$ \\ 
\hline \hline 
$D<0.4$ Mpc	& $-$1.45$\pm$0.01	\\ 
$0.4<D<0.7$ Mpc	& $-$1.27$\pm$0.01	\\ 
$0.7<D<1$ Mpc	& $-$1.29$\pm$0.01	\\ 
$1<D<1.5$ Mpc	& $-$1.21$\pm$0.01	\\ 
$D>1.5$	Mpc	& $-$1.10$\pm$0.03	\\ 
\hline 
\end{tabular} 
\end{table} 

\section{Orientation} 

\begin{figure}[h!] 
\centering 
\includegraphics[width=1\columnwidth]{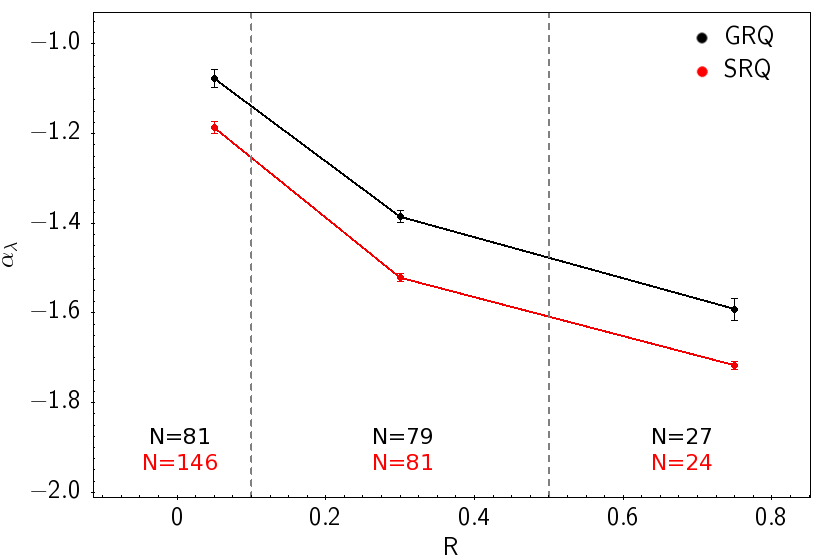} 
\caption{The power law slope vs. core-to-lobe ratio parammeter (R).} 
\label{R_slope} 
\end{figure}

Another parameter that is considered to be related to the slope of the spectral continuum is the orientation of an AGN. Both the dusty torus, as well as an accretion disk, lead to orientation-dependent emission visible in the optical spectrum (e.g., \citealt{nenkova2008, baker1997}). According to commonly accepted theories, the continuum emission increases as the angle between the jet axis of AGN and the line of sight decreases. However, precise determination of the radio source orientation is difficult and it requires the highest resolution radio observations \citep{ghisellini1993}. Usually, in literature, the core-to-lobe ratio (R) is used as an approximation of radio source orientation (e.g., \citealt{kapahi1982}). It is defined as the ratio between the core flux density and the extended radio emission flux density. A high value of the R parameter indicates that the radio source jets are oriented closer to the line of sight. For the samples of GRQs and SRQs, we check the dependence of spectral slope and core-to-lobe ratio using R estimations from \cite{kuzmicz2021}. We binned the quasars according to the R value and computed the composite spectra in each bin. In Figure \ref{R_slope} we plot $\alpha_{\lambda}$ for GRQs and SRQs and list them in Table \ref{tab3}. There is a clear dependence of the steepening continuum for higher R values, which is in agreement with the theoretical predictions and results obtained by, e.g., \cite{becker1995}. \\ 
The other possibility for the determination of radio source orientation is the calculation of the inclination angle of radio lobes, assuming that the Doppler boosting is the main reason of flux density asymmetry between opposite radio lobes. However, for Mpc-scale radio lobes, such orientation estimation can be very different from the orientation of the accretion disc, therefore this method is a very inaccurate way to estimate the orientation of most the central parts of AGN from which we observe optical continuum. To confirm this, we compared the inclination angles for GRQs (taken from \citealt{kuzmicz2012}) with the core-to-lobe parameter and we did not obtain any significant correlation between those two orientation indicators.    

\begin{table}[h!] 
\centering 
\caption{Slopes of the composite continuum for different core-to-lobe ratio (R) bins obtained for GRQ and SRQ samples.} 
\label{tab3} 
\begin{tabular}{ccc} 
\hline 
bin			& $\alpha_{\lambda}$ for GRQ & $\alpha_{\lambda}$ for SRQ \\ 
\hline \hline 
$R<0.1$		& $-$1.08$\pm$0.02	& $-$1.19$\pm$0.01\\ 
$0.1<R<0.5$	& $-$1.39$\pm$0.01	& $-$1.52$\pm$0.01\\ 
$R>0.5$		& $-$1.59$\pm$0.02	& $-$1.72$\pm$0.01\\ 
\hline 
\end{tabular} 
\end{table} 

\section{Spectral energy distribution for individual giant radio quasars} 

\begin{figure}[h!] 
\centering 
\includegraphics[width=1.1\columnwidth]{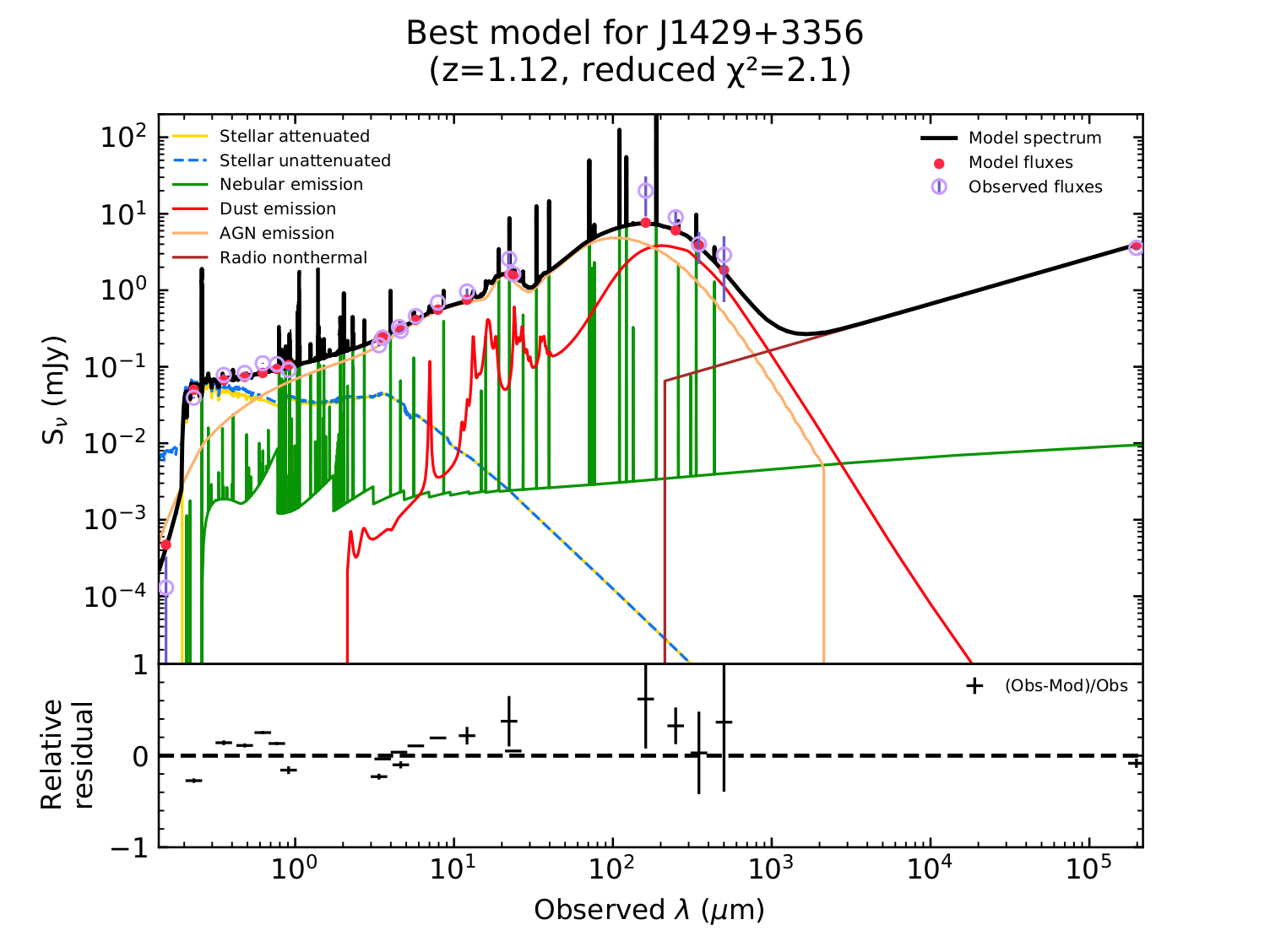}\\ 
\includegraphics[width=1.1\columnwidth]{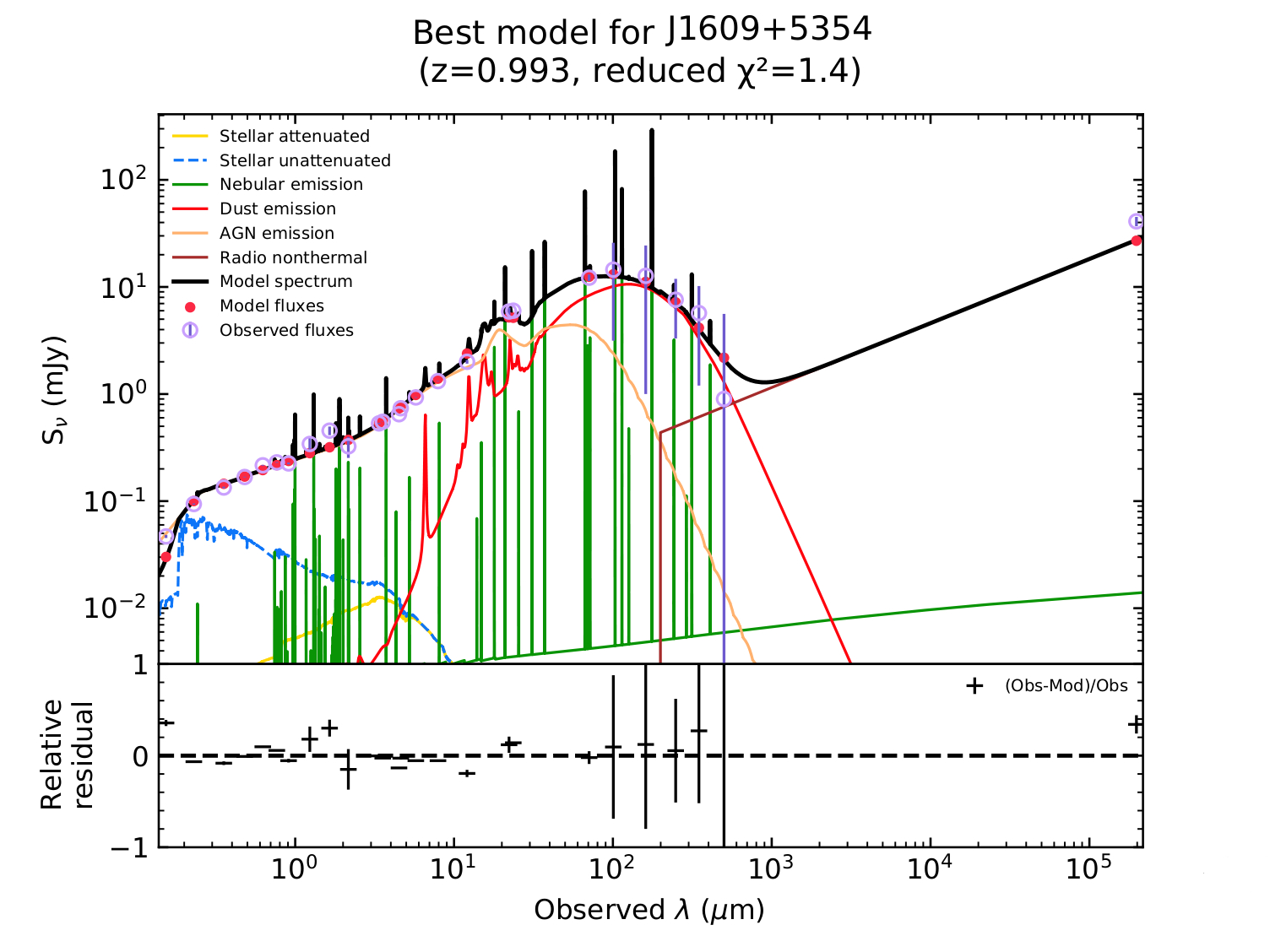}
\caption{Best SED fitting model for the GRQ J1429+3356 and J1609+5354 plotted with the black solid line. Observed fluxes are plotted with open violet circles and the model fluxes are plotted with filled red circles. The best fit is a superposition of AGN emission (orange line), dust-attenuated stellar emission (yellow line; the intrinsic stellar emission is shown by dashed blue line), dust emission (red line) and nebular emission (green line). The relative residual fluxes are plotted under each spectrum.} 
\label{sed} 
\end{figure}

The UV/optical spectral continuum is a small part of the broadband SED of quasars. The studies of SEDs are the key tool in understanding the nature of the physical processes taking place in AGNs, therefore they are very useful in testing theoretical models, as well as in understanding the AGN/host galaxy coevolution. However, for the analysis of broad-band SEDs we need multifrequency observations. As the SED modelling includes electromagnetic bands from UV to the radio. Therefore the more the number of observations is available for a source, the more accurate fit can be obtained. Unfortunately, for the majority of GRQs there are no sufficient data available, therefore only for a few GRQs the SED modelling was already done (e.g., \citealt{colafrancesco2016, hernandez2017}). In our sample we identified two GRQs, J1429$+$3356 ($\alpha$=14$^{\rm h}$29$^{\rm m}$42.65$^{\rm s}$ $\delta$=$+$33$^{\circ}$56$^{\prime}$54.8$^{\prime\prime}$, z=1.12) and J1609$+$5354 ($\alpha$=16$^{\rm h}$09$^{\rm m}$13.19$^{\rm s}$ $\delta$=$+$53$^{\circ}$54$^{\prime}$29.7$^{\prime\prime}$, z=0.99) with near and far UV, optical, near, mid and far IR photometry data along with radio at 1.4 GHz data available. For both quasars the SED fitting was performed using a new version of Code Investigating GAlaxy Emission (CIGALE; \citealt{boquien2019}) so-called X-CIGALE code \citep{yang2020}. We used different modules from X-CIGALE which correspond to a unique physical component or process taking place in an AGN host. The input parameters used for different X-CIGALE modules are listed in Table \ref{input}. The best SED fits for the two GRQs are ploted in Figure \ref{sed}. We obtained the reduced $\chi^2$=2.1 and $\chi^2$=1.4 for J1429$+$3356 and J1609$+$5354 respectively. 

\begin{table}[h!] 
\centering 
\caption{List of input parameters of the X-CIGALE code.} 
\label{input} 
\begin{tabular}{p{0.6\linewidth} p{0.3\linewidth}} 
\hline 
parameter	& value \\ 
\hline \hline 
\multicolumn{2}{c}{Star formation history}\\
\multicolumn{2}{c}{(\texttt{sfhdelayed} module)}\\
\hline
$e$-folding time of the main stellar population, $\tau$ (Myr)	&	3000, 4000\\
Stellar age, $t$ (Myr)	&	3000, 4000, 5000, 6000, 10000\\
Mass fraction of the late burst population &	0.1, 0.2, 0.3\\
\hline
\multicolumn{2}{c}{Single stellar population}\\
\multicolumn{2}{c}{(\texttt{bc03} module; \citealt{bruzual2003})}\\
\hline
Initial mass function	&	\citealt{chabrier2003}\\
Metallicity	& 0.02	\\
\hline
\multicolumn{2}{c}{Attenuation curve }\\
\multicolumn{2}{c}{(\texttt{dustatt\_modified\_starburst} module; }\\
\multicolumn{2}{c}{\citealt{calzetti2000})}\\
\hline
The colour excess of the nebular lines light	&	0.05, 0.1, 0.2, 0.3, 0.4, 0.5\\
E(B$-$V) factor	&	0.1, 0.2, 0.3, 0.4, 0.5, 0.6\\
\hline
\multicolumn{2}{c}{Dust emission }\\
\multicolumn{2}{c}{(\texttt{dl2014 module}; \citealt{draine2014})}\\
\hline
IR power-law slope $\alpha$	&	1.5, 2, 2.5, 3\\
\hline
\multicolumn{2}{c}{AGN emission}\\
\multicolumn{2}{c}{(\texttt{skirtor2016} module; \citealt{stalevski2012, stalevski2016})}\\
\hline
Ratio of outer to inner torus radius	&	10, 20\\
Inclination angle $i$ (face on: $i=0^{\circ}$)	&	10, 20, 30, 40\\
AGN fraction in total IR luminosity	&	0.5, 0.6, 0.7, 0.8, 0.9\\
E(B$-$V) of polar dust	&	0.001, 0.01, 0.1, 0.2, 0.3\\
\hline 
\end{tabular} 
\end{table} 

The main physical parameters of GRQ's hosts are listed in Table \ref{param}. As the GRQ J1429$+$3356 and J1609$+$5354 have steep optical continua with $\alpha_{\lambda}=-1.96$ and $\alpha_{\lambda}=-1.77$ respectively, it is expected that the reddening caused be dust or orientation effects should by very small in those two GRQs. The luminosity absorbed by the dust (dust luminosity in Table \ref{param}) is smaller for J1429$+$3356, what is in consistence with lower value of $\alpha_{\lambda}$. Also, both GRQs have small inclination angles resulting in less dust contamination of the central AGN. We compared the modelled inclination angles with lobe-to-core ratio and obtained the smaller value of inclination for the higher R parameter, as it is expected. \\

\begin{table}[h!] 
\centering 
\vspace{0.5cm}
\caption{The physical parameters for GRQ J1429$+$3356 and J1609$+$5354 obtained through X-CIGALE SED modelling.} 
\label{param} 
\begin{tabular}{lcc} 
\hline 
parameter & J1429$+$3356 & J1609$+$5354 \\ 
\hline \hline 
AGN luminosity & 8.1$\times 10^{38}$ W& 2.5$\times 10^{39}$ W\\
Dust luminosity & 3.4$\times 10^{38}$ W& 4.3$\times 10^{38}$ W\\
Disk luminosity & 14.7$\times 10^{38}$ W& 2.1$\times 10^{39}$ W\\
Stellar luminosity & 6.1$\times 10^{38}$ W& 5.2$\times 10^{38}$ W\\
Total stellar mass & 2.2$\times 10^{10}$ M$_{\odot}$& 1.5 $\times 10^{10} $M$_{\odot}$\\
Total gas mass & 1.2$\times 10^{10}$ M$_{\odot}$& 8.6$\times 10^{9}$ M$_{\odot}$\\
Inclination angle & 18.8$^{\circ}$ & 11.7$^{\circ}$\\
AGN fraction contributing & 0.8 & 0.5\\
to the total IR luminosity&&\\
\hline 
\end{tabular} 
\end{table} 

\section{Conclusions}
In this paper, we study the shape of the optical continuum for a sample of GRQs, as well as SRQs. We compiled the composite spectra for both samples and obtained continuum slopes $\alpha_{\lambda}=-1.25\pm 0.01$ and $\alpha_{\lambda}=-1.34\pm 0.01$ for GRQs and SRQs respectively. The observed continuum slopes are flatter (the spectra are redder) compared to the composite spectrum for SDSS quasars studied by \cite{vanden2001} where the authors obtained $\alpha_{\lambda}=-1.56$. The redder spectra can be a result of intrinsic differences in dust properties or its distribution in the centrer of AGN, but aslo the orientation effects can play a significant role in reddening the spectra.

We also find that the quasar continuum becomes steeper for higher total and core radio luminosities at 1.4 GHz. It confirms the scenario where the radio jet emission is closely connected with accretion disk and central black hole properties. The UV/optical emission generated in accretion processes produces the steeper continuum for higher black hole spins which are believed to be responsible for radio jet production. However, according to theoretical modelling and observational results, the mechanism of radio jet generation in GRQs and other radio sources must be much more complicated.\\
We show that optical spectra become flatter for larger projected linear size of radio structures, which can be closely related with the $\alpha_{\lambda}$ -- $\log$P$_{\rm tot}$ dependence. Thus, according to radio source evolutionary models, it can be interpreted as flattening the spectrum with the age of the radio source.

For GRQs and SRQs we obtained that UV/optical continuum becomes steeper for higher core-to-lobe ratios confirming the dependence between $\alpha_{\lambda}$ and radio source orientation. 

The results obtained from the SED fitting for J1429$+$3356 and J1609$+$5354 are consistent with results obtained from optical continuum slope analysis. However, the small number of GRQs with multiwavelength data available, do not allow to draw more general conclusions concerning GRQs.

\section{Acknowledgments}

We thank Katarzyna Ma{\l}ek for her valuable comments and help in SED modelling.\\
This paper was supported by the National Science Centre, Poland through the grant 2018/29/B/ST9/01793.





\end{document}